\documentstyle[12pt]{article}
\renewcommand{\baselinestretch}[1]{0.23in}
\newcounter{dump}
\begin{document}
\begin{titlepage}
\fnsymbol{footnote}
{\center{\Huge Nearby CFT's in the operator formalism:\\ The role of a
connection\\}}
{}~\\

{\center{\Large K. Ranganathan \footnote[1]{Supported in part by funds provided
by the US Department of Energy(DOE) under contract \# DE-AC02-76ER03069.}\\}}

{\center{\small {\em Center for Theoretical Physics,\\ Laboratory for Nuclear
Science\\ and Department of Physics, \\Massachusetts Institute of Technology,\\
Cambridge, Massachusetts 02139, USA\\~\\~\\~\\~\\}}}

\begin{normalsize}
\begin{abstract}
There are two methods to study families of conformal theories in the operator
formalism. In the first method we begin with a theory and a family of deformed
theories is defined in the state space of the original theory.
In the other there is a distinct state space for each theory in
the family, with the collection of spaces forming a vector
bundle. This paper establishes the equivalence of a deformed
theory with that in a nearby state space in the bundle via a connection that
defines maps between nearby state spaces. We find that an appropriate
connection for establishing equivalence is one that arose in a recent
paper by Kugo and Zwiebach. We discuss the affine geometry induced on the
space of backgrounds by this connection.  This geometry is the same as the
one obtained from the Zamolodchikov metric.
\end{abstract}
\end{normalsize}
{}~\\~\\~\\

{\center {\em Submitted for publication to}  Nuclear Physics B \\}

{}~\\~\\
\noindent CTP\#2154\hfill October 1992
\end{titlepage}

\section{Introduction}
\par
\par The basic problem facing string theory is that we know no principle to
select the right background (i.e the right Conformal Field theory, henceforth
CFT) from the plethora of consistent string models. One line of work, places
faith in String Field Theory(SFT) as a route to discovering a formulation with
more predictive power than the present one. By investigating the relation
between the SFT actions of ``nearby" backgrounds, one hopes to progress toward
an analogue of SFT, that does not require choosing of a specific background.
Since the SFT action is constructed using the operator formalism it is natural
in these investigations to use this formalism.

\par{\em \underline{Deforming CFT's:}} In the operator formalism two methods
for the study of nearby CFT's, have received attention. One method formulates
the new CFT in the state space of the old one. An explanation of this statement
follows. CFT's are, leaving aside subtleties, representations of the algebra of
${\cal P} (g,n)$, the space of $n$ punctured Riemann surfaces with coordinates
chosen around the punctures. This means that for every point $p \in {\cal
P}(g,n)$ we have an $n$ tensor, $|p\!>$ in the space of tensors $T$ on the
state space $H$ and that the algebraic operation of sewing surfaces in ${\cal
P} (g,n)$ is realized as an operation on the associated tensors. By saying that
the new CFT is formulated in the state space of the old we mean that for each
point $p \in {\cal P}(g,n)$ we have a new $n$ tensor $|p\!>^{new}$ defined on
the same state space such that the CFT axioms (essentially the sewing
requirement) hold. In practice we do not know how to define the $|p\!>^{new}$
exactly. Instead one assumes that the new theory is defined in a power series
in a small parameter $\epsilon$, with an explicit prescription given only for
the first two terms. This method was developed by many authors \cite{many} and
was summarized succinctly by Nelson, Campbell and Wong \cite{NCW}.  Sen
\cite{S} has used this method to study the relation between SFT actions of
theories related by ``exactly marginal" perturbation.

\par{\em \underline{Nearby CFT's in distinct state spaces:}} Kugo and Zwiebach
\cite{KZ} discuss another approach to exploring nearby conformal theories. They
begin with a specific space of backgrounds with a different state space for
each background.  The space of backgrounds $B$ they consider, arises through
toroidal compactification. A specific background is characterized by a matrix
$E$ of dimension $d$ equal to the number of compactified dimensions. This
matrix contains the compactification data \cite{NSW}. The state space of the
background at $E$ is denoted $H_E$ and is constructed through the action of
oscillators $\alpha_m^{(i)}(E)$ and $\bar{\alpha}_m^{(i)}(E)$ on the vacum
$|0\!>_E$, for all $m > 0$ and $i = 1...d$. In mathematical language then, what
we have is a vector bundle $V_B$ with base space $B$ and fibers $H_E$. One then
defines a ``connection", $\Gamma_{KZ}$, on the vector bundle $V_B$ to
infinitesimally transport arbitrary tensors - and hence those that define CFT
correlation functions - from one background to another (i.e from one basepoint
to another).

\par{\em \underline{The result of this paper:}} The method described by Nelson,
Campbell and Wong in \cite{NCW} is generally applicable to any CFT. It can in
particular be applied to the background $E$. Doing so we obtain a new theory
defined in $H_E$. We expect this new theory to be equivalent to the theory
defined at a background $E'$, for some $E'$. This derives from the idea that
deforming theories a la NCW ``corresponds" to adding a term to the
Lagrangian in functional integration and hence that the deformed theory and the
theory at $E'$ are different operator realizations of the same 2-d lagrangian.
To our knowledge this plausibility argument for equivalence has not yet been
turned into a proof within the operator formulation of CFT's. The main result
of this paper is the proof of the equivalence of a deformed theories with that
at a different background to first order in the perturbing parameter
$\epsilon$. This result can be restated as follows. We establish the existence
of a connection on the vector bundle $V_B$ with the property that the parallel
transport it defines (to first order in $\epsilon$) establishes the equivalence
of a deformed theory in $H_E$ to the theory defined in $E'$. We note that in
another context, that of quatum field theories, Sonoda \cite{S} has dealt with
connections on the vector bundle of theories.

We will now explain the meaning of this result. First we detail what it means
for two CFT's to be equivalent in the operator formalism. Equivalence means
that there is an isomorphism (a one to one and onto map), $f_{EE'}$ between
$H_{E'}$ and $H_{E}$ with the property that the map $f^T_{EE'}$ it induces
between the space of tensors $T_{E'}$ and $T_{E}$, carries the tensors,
$|p\!>_{E'}$, of the theory at $E'$ into the corresponding tensors
$|p\!>_{E}^{new}$, of the new theory at $E$. In terms of the diagram of Fig. 1
below we require that the maps in the triangle commute (i.e $CFT_{new} =
f^T_{EE'} \cdot CFT$). The maps named $CFT$ from ${\cal P}(g,n)$ to $T_E$ and
$T_{E'}$ are the ones that define the $CFT$ in the respective backgrounds
whereas $CFT_{new}$ is the one that defines the deformed theory in the state
space $H_E$. We will require in addition that the map $f_{EE'}$ carry the
complex conjugation operator, $C_E$ in one space into the conjugation operator
$C_{E'}$ in the other. Recall that the operator $C_E$ together with the BPZ
product defines the hermitian inner product on the state space. Our condition
on $C_E$ then implies that $f_{EE'}$ preserves the hermitian inner product.

\begin{picture}(400,230)
\put (100,20){\makebox(0,0){$H_E$}}
\put (300,20){\makebox(0,0){$H_{E'}$}}
\put (265,20){\vector(-1,0){145}}
\put (200,30){\makebox(0,0){$f_{EE'}$}}
\put (300,35){\line(0,1){50}}
\put (100,35){\line(0,1){50}}
\put (100,100){\makebox(0,0){$T_E$}}
\put (300,100){\makebox(0,0){$T_{E'}$}}
\put (265,100){\vector(-1,0){145}}
\put (200,110){\makebox(0,0){$f^T_{EE'}$}}
\put (210,195){\vector(1,-1){80}}
\put (270,155){\makebox(0,0){$CFT$}}
\put (190,195){\vector(-1,-1){80}}
\put (175,155){\makebox(0,0){$CFT_{new}$}}
\put (175,130){\oval(150,150)[tl]}
\put (100,132){\vector(0,-1){15}}
\put (80,165){\makebox(0,0){$CFT$}}
\put (200,205){\makebox(0,0){${\cal P}(g,n)$}}
\put (200,-10){\makebox(0,0){{\em Figure 1}}}
\end{picture}

{}~\\

An isomorphism $f_{EE'}$ that makes the triangle of maps commute is not unique
since there could be automorphisms of the state space which preserve all the
states $|p\!>$ of the CFT - that is there could be symmetries. If $S$ denotes a
symmetry of the perturbed CFT at $E$  (i.e $S|p\!>^{new}_E = |p\!>^{new}_E$)
then $S \cdot f_{EE'}$ defines another isomorphism that establishes the
equivalence of theories. The possibility of non trivial maps $S$ (not equal to
the identity) exists because the set of tensors $|p\!>$ that arise from a CFT
only constitute a subset (not a subspace) of the space of all tensors. A non
trivial symmetry $S$ preserves the tensors in this subset but must change some
of the tensors outside of this subset.

Having detailed the meaning of equivalence of conformal theories in the
operator formalism we define what it means for a deformed theory to be
equivalent to a nearby one to the first order in the perturbing parameter
$\epsilon$. To do so we introduce a connection $\Gamma$ on the vector bundle
$V_B$. A connection, $\Gamma$ associates with every path between $E'$ and $E$ a
map from $H_{E'}$ to $H_{E}$, and hence a method of transporting tensors. In
particular the connection can be used to transport the tensors arising from the
CFT map. Instead of paths we will be considering the case $E'$ being nearby to
E, that is $E' = E + \epsilon \delta E$, where $\delta E$ is thought of as an
element of the tangent space $T_E B$. The pair $E$ and $E'$ is all that is
necessary to define the transport of tensors to first order in $\epsilon$. The
transported tensor in $T_E$ is denoted $\Gamma(E,E')|p\!>_{E'}$. Here
$\Gamma(E,E')|p\!>_{E'}$ is understood to include a factor of $\epsilon$.  Note
that this first order transport does not require us to solve the equation of
parallel transport for a connection. Now we need to say to which deformed
theory at $E$ we are going to compare the theory transported from $E'$.
To answer this we recognize that the perturbing operators, $\partial_z x^i
\delta E_{ij} \partial_{\bar{z}} x^j$, that define the deformation of a theory
a la \cite{NCW} (see Eqns.(~\ref{mainncw}),(~\ref{phi})) are defined by a
matrix $\delta E$, of the same dimension as the one that defines a specific
background. There is then a natural identification of $T_E B$, the tangent
space to $B$ at the point $E$ with the subspace of ``exactly marginal" states
in $H_E$, denoted $M_E$. This identification defines the deformed theory to
which we are to compare the theory transported from $E'$. If the transported
tensors are to be equal to the deformed tensors to first order in $\epsilon$
then the connection must satisfy the equation below
\begin{equation}
\Delta |p\!>_E = \Gamma (E,E')|p\!>_{E'} - |p\!>_E \;\; \forall p \in {\cal
P}(g,n), \label{gcfe}
\end{equation}

\noindent where $\Delta |p\!>_E $ denotes the first order correction in
deforming theories by the method described in \cite{NCW}(See
Eqn.(~\ref{mainncw})). In this paper we will show the existence of a connection
satisfying this condition. This connection was first defined in \cite{KZ} and
we denote it $\Gamma_{KZ}$. If we define
\begin{equation}
\delta |p\!>_E = \Gamma_{KZ}(E,E')|p\!>_{E'} - |p\!>_E.
\end{equation}

\noindent then the condition of Eqn.(~\ref{gcfe}) can be written as
\begin{equation}
 \Delta |p\!>_E  = \delta |p\!>_E \;\; \forall p \in {\cal
P}(g,n)\;\;\;\;\;\;\;\;\;\;\;\;\mbox{\underline{The condition for first order
equivalence}} \label{cfe}
\end{equation}

\noindent Verifying this equation, henceforth called the condition for first
order equivalence, is the main task of this paper.

\par{\em \underline{Geometry induced by connection:}} Having noted that the
identification between the tangent space $T_E B$ and the space of exactly
marginal states $M_E$ is not without significance we investigate the geometry
of $B$. The identification suggests a natural candidate for a metric on $B$,
the Zamolodchikov metric.We will denote the connection associated with this
metric as $\Gamma_Z$. This connection on the space of theories is related to
the contact terms that arise in conformal theory and was investigated by
Kutasov \cite{K}. Note that this is a connection on the manifold
$B$ as distinguished from a connection on the vector bundle $V_B$, such as
$\Gamma_{KZ}$. However $\Gamma_{KZ}$ gives rise to another
candidate for a connection on $B$. When  $\Gamma_{KZ}$ is used to transport
states in $M_{E'}$ to $H_{E}$, the transported state will not in
general lie in
$M_{E}$. If however it is projected back into $M_{E}$ with the BPZ metric, a
connection $P\Gamma_{KZ}$ on the space $B$ is obtained through the
identification of $T_E B$ and $M_E$. We will show that the two connections,
$\Gamma_Z$ and $P\Gamma_{KZ}$ are the same.

\par{\em \underline{Issues not addressed in this paper:}} There are other
issues that still need investigation with regard to connections that satisfy
the condition for first order equivalence Eqn.(~\ref{cfe}). Firstly we would
like to establish the curvature of this connection in the vector bundle. If it
is flat then the map that it defines between points $E'$ and $E$ are
independent of path. If not we have a family of maps that identify a deformed
theory with that at a different point.  Secondly the remarks on the
non-uniqueness of the map $f_{EE'}$ suggest that there is no unique connection
satisfying Eqn.(~\ref{cfe}). What then is the space of such connections ?
Thirdly if the condition of flatness is imposed on the connection what subset
of this space are we restricted to ? Lastly for a connection the holonomy
around any closed curve would yield symmetries of the theory at a particular
background and this would be interesting to study as well.

The investigation of these issues requires that we resolve the problems
associated with higher order corrections both in deforming conformal theories
and in defining parallel transport with the connection. In our present
understanding the higher order corrections in deforming conformal theories are
divergent. In addition the prescription for transporting states from one
background to another also yields divergences if we attempt to include higher
order corrections. In this paper we have not had to deal with these problems
since we work not with actual deformations or actual maps, but only with their
infinitesimally analogues. While the condition for first order equivalence,
Eqn.(~\ref{cfe}), is a necessary requirement on a connection that establishes
equivalence of theories, it is not sufficient. Unless we show that actual maps
can be obtained through parallel transport and that actual deformations exist
we would not have established the equivalence of deformed theories to those in
a neighbouring background. We hope to address the issue of higher order
corrections in the near future.

\par{\em \underline{Plan of this paper:}} This paper is organized as follows.
The task of sections 2 through 6 is the verification of the condition for first
order equivalence, Eqn.(~\ref{cfe}). Section 2 is a brief review of the
standard method of deforming theories which was summarized in \cite{NCW}.
Section 3 reviews the necessary parts of the \cite{KZ} paper. In Section 4 we
argue that in order to establish the Eqn.(~\ref{cfe}) it is sufficient to (a)
check that $\Delta|p\!>_E = \delta|p\!>_E$ for one point in each of the spaces
${\cal P}(0,1)$, ${\cal P}(0,2)$ and ${\cal P}(0,3)$ and (b) check that $\Delta
L_n(E) = \delta L_n(E)$ where $L_n$ are the Virasoro operators (To understand
this equation recall that an operator such as $L_n$ is also a tensor). This
argument uses the sewing property of tensors in conformal field theory and the
transport properties of the Virasoro operators. In the appendix we show that
Eqn.(~\ref{cfe}) is satisfied for all contact interactions. A contact
interaction is one where the $|\xi (z)|=1$ discs of the local coordinates
$\xi(z)$ exactly cover the entire surface. This result ensures that we have at
least one point each in  ${\cal P}(0,2)$ and ${\cal P}(0,3)$ with the requisite
property. This however leaves out the case of ${\cal P} (0,1)$ where the notion
of contact interaction is not meaningful. Section 5 remedies this by showing
that $\Delta |S_{\frac{1}{z}}\!> = \delta |S_{\frac{1}{z}}\!>$ where
$|S_{\frac{1}{z}}\!>$ is the state associated to the sphere with one puncture
at $\infty$ with standard coordinate $1/z$. In section 6 we show that $\Delta
L_n(E) = \delta L_n(E)$ where $L_n$ are the Virasoro operators. This completes
the proof of the condition for first order equivalence. In section 7 we
consider (affine) geometry on the space of backgrounds $B$ and establish that
$\Gamma_Z = P\Gamma_{KZ}$.  Section 8 is devoted to conclusions and questions
that remain to be answered.

\section{Review of Ref {[NCW]}}

\par The paper of Nelson, Campbell and Wong \cite{NCW} describes how, given a
CFT one can define a new one in the same Hilbert space. More precisely we think
of the tensors of the new CFT as being defined by a power series in a parameter
$\epsilon$. The zero order term for the tensor associated to a point $p \in
{\cal P}(g,n)$ is the original tensor, and the first order term is obtained by
summing over (integrating) contributions from tensors associated with points in
${\cal P}(g,n+1)$. More concretely we write equation 3.1.1 of \cite{NCW}
\begin{equation}
 |\Sigma_R, \xi \! >_{\epsilon} \: \equiv \: |\Sigma_R, \xi \! >_0 + \:
\overbrace{\frac{\epsilon}{2 \pi i} \int_{P \in \Sigma_R/D_{\xi}}  < \!
\phi|\Sigma_R, \xi, u \! >_0 \, du|_P \wedge d\bar{u}|_P}^{\Delta |p\!> },
\label{mainncw}
\end{equation}

\noindent where $\Sigma_R$ denotes the Riemann surface with punctures, $\xi$
denotes the degrees of freedom corresponding to the location of the punctures
and the local coordinates chosen around the punctures. Together they determine
the point $p \in {\cal P}(g,n)$. The tensor corresponding to this point is
denoted, in the original theory as $|\Sigma_R, \xi \! >_0$ and in the deformed
theory as $|\Sigma_R, \xi \! >_{\epsilon}$. The bra $<\!\phi|$ is the BPZ
partner of $|\phi\!>$, a state which is primary and of dimension (1,1). $P$ is
the point at which this state is inserted in the n+1 point function. The
surface $\Sigma_R$ is covered by a set of coordinate patches of which $u$ is a
representative. The local coordinate chosen at the point $P$ is $u - u(P)$.
$D_{\xi}$ denotes the union of interiors of the discs $\xi_i \leq 1$
corresponding to the punctures $i$. The domain of integration of the point $P$,
is the exterior of the discs $|\xi_i(z)| \leq 1$ on $\Sigma_R$.

The prescription of Eqn.(~\ref{mainncw}) for $\Delta |p\!>$ ensures that the
sewing property required of a conformal theory holds. If the deformed tensors
are sewn together then the first order term in $\epsilon$ is clearly just the
integral over the sewn surface of the insertion of the state $|\phi\!>$. This
is precisely the statement that the sewing property is preserved to first
order. The prescription of Eqn.(~\ref{mainncw}) works only if $|\phi\!>$ is
primary and of dimension (1,1) since otherwise the integral of the insertion
over the surface will depend on the coordinate patches chosen to evaluate the
integral.

In this paper we will be applying this method to the specific space of
backgrounds described in the next section.

\section{Review of Ref {[KZ]}}
The space of backgrounds we consider in this paper is a space of toroidally
compactified backgrounds, $B$. More details on this space of theories and
elaboration on the discussion below may be found in \cite {KZ}. A point $b \in
B$ is characterized by a matrix $E$, which contains the information on the
metric and the antisymmetric tensor in the compactification. Associated with
every point $b \in B$ there is a state space $H_b$. For every point in ${\cal
P}(g,n) {\tt x} B$ we have an n tensor defined in the associated state space,
$H_b$, as required by the axioms of conformal theory. In other words, for every
$p \in {\cal P}(g,n)$ we have a field of tensors $|p\!>_E$ on the space $B$.

Consider backgrounds $E$ and $E'$ where $E' = E + \epsilon \delta E$ with
$\epsilon$ a small parameter and $\delta E$ a matrix. Consider a map between
the state space of the backgrounds $E'$ and $E$ which is given as a power
series in the small parameter $\epsilon$.  This allows us to transport tensors
from the background $E'$ to the background $E$. The transported tensors are
also represented as a power series in $\epsilon$. If however we are given only
the the zeroth and first order terms of the map then we can compute the
transported tensors to first order. We will refer to tensors transported in
this way as ``tensors transported by the connection".

How is such a connection defined ? To answer this we need to look at how the
theory at each background is constructed. The construction involves quantizing
a 2-d action of the fields $X^i(\sigma,\tau)$. The construction of the theory
through canonical quantization of the appropriate Hamiltonian involves
expanding the fields $X^i(\sigma)$ and the momentum associated with it,
$P_i(\sigma)$, in terms of oscillators $\alpha_n^i$ as follows
\begin{eqnarray}
X^i(\sigma)|_E & = & x^i + w^i + \frac{i}{\surd
2}\sum_{n\neq0}\frac{1}{n}{[\alpha_n^i(E)e^{in\sigma} +
\bar{\alpha}_n^i(E)e^{-in\sigma}]}, \\
2 \pi P_i(\sigma)|_E & = & p_i + \frac{1}{\surd
2}\sum_{n\neq0}{[E^t_{ij}\alpha_n^j(E)e^{in\sigma} + E_{ij}
\bar{\alpha}_n^j(E)e^{-in\sigma}]},
\end{eqnarray}

\noindent where
\begin{equation}
{[x^i,p_j]} = i\delta_j^i,
\end{equation}
\noindent and
\begin{equation}
{[{\alpha}_m^i(E), {\alpha}_n^j(E)]} = {[\bar{\alpha}_m^i(E),
\bar{\alpha}_n^j(E)]} = mG^{ij}\delta_{m+n,0}.
\end{equation}

\noindent To obtain how the oscillators at a background E are to be mapped to
those at a background $E'$ the $X^i(\sigma)$ and $P_i(\sigma)$ are regarded as
``universal coordinates" on the space of backgrounds. In equations this means
\begin{equation}
X^i(\sigma)|_E \equiv X^i(\sigma)|_{E'} \mbox{	and }  P_i(\sigma)|_E =
P_i(\sigma)|_{E'}.
\end{equation}
\noindent  The mapping of the oscillators is in itself not sufficient to define
a map between the state spaces at different points. To obtain a candidate for a
map we would in addition have to map one state in $H_{E'}$ to $H_{E}$. We can
then generate the full state space by the action of the oscillators on this
state. Not surprisingly the state chosen is the vacuum state. The relation
between fock space states and wave functionals suggests that we map the vacuum
as follows
\begin{equation}
|0>_{E'} \rightarrow e^{\cal B} |0>_E \label{bog},
\end{equation}

\noindent where the antihermitean operator ${\cal B}$ generates the Bogoliubov
transformation $e^{\cal B}$ to go from states in $H_{E'}$ to that in $H_{E}$.
This transformation preserves the condition $\alpha_n(E) |0\!>_E = 0 \;\;
\forall n\geq0$ (i.e the transported annihilation oscillators acting on the
transported vacuum is zero). We should note that Eqn. (~\ref{bog}) is formal
since infinities that arise in evaluating the exponential make it ill defined.

However the infinitesimally analogues of these operations can be defined and
this is all we need to obtain the connection mentioned earlier. More
specifically consider a tensor that is described in terms of the action of
oscillators on the vacuum of the theory at $E'$. In equations
\begin{equation}
| V\! >_{E'}  = f_{E'}(\alpha(E'), \bar{\alpha}(E')) |0\!>_{E'}.
\end{equation}

\noindent Now make the following replacements in the above equation and only
retain the terms which are first order in $\epsilon$ to get $\Gamma_{KZ}
(E,E')|V\!>_{E'}$ the tensor in the space $H_{E}$ obtained by transporting with
the connection
\begin{eqnarray}
\alpha_n(E') & \rightarrow & \alpha_n(E) - \frac{\epsilon}{2}G^{-1}(\delta
E^t\alpha_n + \delta E \bar{\alpha}_{-n}), \nonumber\\
\bar{\alpha}_n(E') & \rightarrow & \bar{\alpha}_n(E) - \frac{\epsilon}{2}
G^{-1}(\delta E \bar{\alpha}_n + \delta E^t \alpha_{-n}), \label{conkz}\\
|0>_{E'} & \rightarrow & |0>_E + {\cal B} |0>_E, \nonumber
\end{eqnarray}

\noindent where
\begin{equation}
{\cal B} =  \sum_{p\neq 0}  \frac{1}{2p} \alpha_p^i \delta E_{ij}
\bar{\alpha}_p^j.
\end{equation}

The operation defining the connection works independently on each index of a
tensor, and this ensures that the operation is tensorial. We must now check
that this recipe is well defined in that if a tensor is represented in two
different ways - i.e with two different  polynomials $f_{E'}(\alpha(E'),
\bar{\alpha}(E'))$ - then the recipe yields the same tensor when applied to
either of the representations. The tensorial property of the operation ensures
that we need to check this condition only for first rank tensors (i.e vectors).
The first question then is what is the equivalence class of  polynomials
$f_{E'}(\alpha(E'), \bar{\alpha}(E'))$ that correspond to the same operator in
the Hilbert space ? The answer is that all $f_{E'}(\alpha(E'),
\bar{\alpha}(E'))$ related by commutation of the $\alpha 's$ lie in the same
class. The second question is what are the equivalence class of operators that
yield the same state when acting on the vacuum ? To answer this use the
commutation relations to write every operator in normal ordered form, $N\left
(\frac{}{} f_{E'}(\alpha(E'), \bar{\alpha}(E'))\right )$ as a sum of products.
Consider the coefficients of each of the products. Clearly if the coefficients
of the products composed only of creation operators is the same in two $N\left
(\frac{}{} f_{E'}(\alpha(E'), \bar{\alpha}(E'))\right )$, then they have the
same action on the vacuum.  This understanding of the equivalence class of
$f_{E'}(\alpha(E'), \bar{\alpha}(E'))$ that lead to the same vector when acting
on the vacuum yields the two sets of conditions that need to be satisfied if
the operation is to define a connection. The first condition is that the
commutation algebra of operators should be preserved when mapping them from
$E'$ to $E$. This means that the algebra of operators on the left and right
hand side of Eqns.(~\ref{conkz}) should be the same to first order in
$\epsilon$. For example this requires that
\begin{eqnarray}
\lefteqn{{[\alpha_n(E'), \bar{\alpha}_n(E')]}} \\
& = & {\left [\alpha_n(E) - \frac{1}{2}G^{-1} \epsilon ( \delta E^t\alpha_n +
\delta E \bar{\alpha}_{-n}), \bar{\alpha}_n(E) - \frac{1}{2}G^{-1}\epsilon
(\delta E \bar{\alpha}_n + \delta E^t \alpha_{-n}) \right ]},\nonumber
\end{eqnarray}
\noindent to first order in $\epsilon$. There are two other such equations one
with both oscillators holomorphic and the other with both anti-holomorphic. All
of these conditions can be shown to hold. The second set of conditions is the
requirement that an annihilation operator be carried into an operator that
annihilates the new vacuum.
\begin{eqnarray}
\alpha_n(E) - \frac{1}{2}G^{-1}(\delta E^t\alpha_n + \delta E
\bar{\alpha}_{-n}) (|0>_E + {\cal B}|0>_E)& = &0 \mbox{     for   } n \geq
0,\nonumber\\
\bar{\alpha}_n(E) - \frac{1}{2} G^{-1}(\delta E \bar{\alpha}_n + \delta E^t
\alpha_{-n})(|0>_E + {\cal B} |0>_E)& = &0 \mbox{     for   } n \geq 0.
\end{eqnarray}

\noindent These conditions are guaranteed by virtue of the property of
Bogoliubov transformations described earlier. So we conclude that we have a
well defined connection.  This connection is defined for any tensor $|
V\!>_{E'}$. It can in particular be applied to the states $|p\!>_{E'}$ that
arise in conformal theory. Doing so we are led to define as was done earlier
\begin{equation}
 \delta |p\!>_E = \Gamma_{KZ} (E,E')|p\!>_{E'} - |p\!>_E.
\end{equation}

We have now defined the two operations $\delta$ and $\Delta$. Both define the
first order corrections of the tensors $|p\!>$ that define a conformal theory.
However, the two operations are in spirit and origin very different.  $\delta
|V\!>_E$ can be defined for an arbitrary field of tensors defined on the space
of backgrounds. We don't need to require that it be the field $|p\!>_{E}$
arising from the conformal theories defined at each of the backgrounds. This
distinguishes it from the NCW method which is a pointwise operation and does
not require a field of objects. Further $\Delta$ defines the first order
perturbation only for tensors $|p\!>_E$ that arise from the CFT map rather than
for arbitrary tensors.

\section{Reducing the condition for first order equivalence }
Having completed reviews and definitions we move to the main task of this paper
the verification of the condition for first order equivalence,
Eqn.(~\ref{cfe}). The aim of this section is to reduce the verification of this
condition to, checking  $\Delta |p\!>_E = \delta |p\!>_E$ for a few simple
surfaces $p \in {\cal P}(g,n)$. We start by showing that it is sufficient to
verify that $\Delta |p\!>_E = \delta |p\!>_E$ for just one point in every space
${\cal P}(g,n)$. The argument goes as follows. Recall that for any $p, q \in
P(g,n)$ we can obtain $|q\!>$ by the application of Virasoro operators on
$|p\!>$ (the transport equation of \cite{V}). For $p$ we choose a point where
we have established that $\Delta |p\!>_E = \delta |p\!>_E$. The expression for
$|q\!>$ in terms of the Virasoro operators also yields the prescription for
computing $\Delta |q\!>_E$ and $\delta |q\!>_E$ in terms of $\Delta L_n$ and
$\delta L_n$ respectively (recall that $\Delta L_n$ and $\delta L_n$ are
defined by regarding the $L_n$ as tensors). The prescriptions for $\Delta
|q\!>_E$ and $\delta |q\!>_E$ are identical except for the replacement of
$\Delta L_n$  by $\delta L_n$. Since as we will later show in section 6 $\Delta
L_n = \delta  L_n$ we have the result required.

Now we use the sewing property to show that it is enough to have one point for
which $\Delta |p\!>_E = \delta |p\!>_E$ in each of ${\cal P}(0,1)$, ${\cal
P}(0,2)$ and ${\cal P}(0,3)$ to ensure that there is one $p$ in each ${\cal
P}(g,n)$ with this property. Start with a surface $r$ obtained by sewing
together surfaces $p$ and $q$. Let $<\!S_{z,1/z}|$ denote the tensor
corresponding to the sewing surface. Then we can write
\begin{equation}
|r\!>_{E'} = _{E'}<\!S_{z,1/z}| \left (\frac{}{} |p\!>_{E'} |q\!>_{E'} \right
).
\end{equation}

\noindent Since the connection acts tensorially this expression for
$|r\!>_{E'}$ tells us that
\begin{equation}
\Gamma_{KZ}(E,E')|r\!>_{E'} = {[_{E'}<\!S_{z,1/z}|\Gamma_{KZ}(E,E')]} \left
(\frac{}{} \Gamma_{KZ} (E,E')|p\!>_{E'}  \Gamma_{KZ} (E,E')|p\!>_{E'} \right ).
\end{equation}

\noindent A corollary of the result in the appendix is that
$_{E'}<\!S_{z,1/z}|\Gamma_{KZ}(E,E') = _{E}<\!S_{z,1/z}|$. We then use the
definition $\delta |p\!>_E = \Gamma_{KZ}(E,E')|p\!>_{E'} - |p\!>_E$ and retain
terms only to first order in $\epsilon$ to find
\begin{equation}
\delta |r\!>_{E} = _{E}<\!S_{z,1/z}| \left (\frac{}{} \delta |p\!>_{E}
|q\!>_{E} + |p\!>_{E} \delta |q\!>_{E} \right ).
\end{equation}

\noindent Now if $\delta |p\!>_E = \Delta |p\!>_E$ and $\delta |q\!>_{E} =
\Delta |q\!>_E$ then we see that
\begin{equation}
\delta |r\!>_{E} = _{E}<\!S_{z,1/z}| \left ( \frac{}{}\Delta |p\!>_{E}
|q\!>_{E} + |p\!>_{E} \Delta |q\!>_{E} \right ) = \Delta |r\!>_{E},
\end{equation}

\noindent where in the last step we have used the fact that the deformation of
CFT's preserves the sewing property to first order in $\epsilon$ (see Section
2). To conclude then we have shown how to obtain a surface $r$ for which
$\Delta |r\!>_E = \delta |r\!>_E$ from two surfaces $p$ and $q$ satisfying the
same property. If then we have at least one surface in each of ${\cal P}(0,1)$,
${\cal P}(0,2)$ and ${\cal P}(0,3)$ satisfying this property, then we can
obtain at least one surface $p$ in each ${\cal P}(g,n)$ satisfying $\Delta
|p\!>_E = \delta |p\!>_E$, by sewing together the former.

We will show in the appendix that $\Delta |p\!>_E = \delta |p\!>_E$ explicitly
for all contact interactions in ${\cal P}(0,n)$. Contact interactions are
surfaces where the $|\xi_i(z)| \leq 1$ discs for each puncture cover the entire
surface exactly. This is in fact more than we need to do since we need only one
point in each of ${\cal P}(0,1)$, ${\cal P}(0,2)$ and ${\cal P}(0,3)$ where
this property holds. That we can explicitly verify for more surfaces than
required provides confirmation of our formal arguments, on transport using
Virasoro operators and sewing.

\section{Showing that $\Delta |S_{\frac{1}{z}}\!> = \delta
|S_{\frac{1}{z}}\!>$}

We now consider the case of the sphere with one puncture at $\infty$ and
coordinate $1/z$. The tensor associated with this point in ${\cal P} (0,1)$ is
denoted $|S_{\frac{1}{z}}\!>$. The formula in Eqn. (~\ref{mainncw}) tells us
how to compute the change in the state associated with this point in ${\cal P}
(0,1)$. This formula involves the tensor corresponding to a two punctured
sphere. A two punctured sphere can be obtained from a three punctured sphere by
sewing a one punctured sphere with standard coordinates onto the third
puncture, so that it deletes that puncture. We will use this fact to write the
formula in Eqn.(~\ref{mainncw}) in terms of the conformal field
$\Phi_{z\bar{z}}$, corresponding to the state $|\phi \!>$. Recall that a
conformal field is defined in terms of a three punctured sphere with the state
$|\phi \!>$ inserted in one of the punctures. The action of the conformal
field, $\Phi_{z\bar{z}}$, on the vacuum $|0\!>_E$ yields a two punctured sphere
with a puncture at $\infty$ with coordinate $1/z$ and a puncture at $z$ with
standard coordinate at which a state $|\phi \!>$ is inserted. Referring to the
prescription of Eqn.(~\ref{mainncw}) we see that this is the right object to be
used in the computation of $\Delta |S_{\frac{1}{z}}\!>$. So we find
\begin{equation}
\Delta |S_{\frac{1}{z}}\!> = \int_{|z| < 1} \! dz d\bar{z} \:\:
\Phi_{z\bar{z}}|0\!>_E.
\end{equation}

\noindent Now change variables from $z$ to $r,\theta$ where $z = re^{i\theta}$
and substitute the explicit expression for $\Phi_{z\bar{z}}$ below
\begin{equation}
\Phi_{z\bar{z}} =  \partial_z x^i \delta E_{ij} \partial_{\bar{z}} x^j \mbox{
 where     }\partial_z x^i = \sum_n \alpha_n^i z^{-n-1} \mbox{  and  }
\partial_{\bar{z}} x^j = \sum_n \bar{\alpha}_n^j \bar{z}^{-n-1} \label{phi}.
\end{equation}

\noindent Using these expansions we get
\begin{eqnarray}
\Delta |S_{\frac{1}{z}}\!> & = & \frac{i}{2 \pi i}\int_0^1 dr \, r \int_0^{2
\pi} d \theta \,\sum_p \alpha_p^i r^{-p-1} e^{-i(p+1)\theta} \delta E_{ij}
\sum_q \bar{\alpha}_q^j r^{-q-1} e^{i (q+1) \theta} |0 \! >, \nonumber\\
& = & \frac{1}{2 \pi }\int_0^1 dr \, r^{-p-q-1}  \sum_p \sum_q \int d\theta \,
e^{i(q-p)\theta}\alpha_p^i \delta E_{ij}  \bar{\alpha}_q^j  |0 \! >,
\nonumber\\
& = & \sum_p \int_0^1  dr \, r^{-2p-1} \: \alpha_p^i \delta E_{ij}
\bar{\alpha}_p^j \: |0 \! >, \nonumber\\
& = & \sum_{p<0} \frac{1}{2p} \alpha_p^i \delta E_{ij}  \bar{\alpha}_p^j
|0\!>.  \label{dncw0}
\end{eqnarray}

\noindent where the last step uses the fact that $\alpha_p^i |0 \! > =
\bar{\alpha}_p^j |0 \! > = 0$ for $p \geq 0$. Having computed $\Delta
|S_{\frac{1}{z}}\!>$ let us compute $\delta |S_{\frac{1}{z}}\!>$. Eqn
(~\ref{conkz}) tells us that
\begin{equation}
\delta |0\!> = -{\cal B} |0\!> \mbox{     where     } {\cal B} =  \sum_{p\neq
0} \, \frac{1}{2p} \, \alpha_p^i \delta E_{ij}  \bar{\alpha}_p^j.
\end{equation}

\noindent Using again $\alpha_p^i |0\!> = \bar{\alpha}_p^j |0\!> = 0$ for $p
\geq 0$ we get
\begin{equation}
\delta |0 \! > = \sum_{p<0} \, \frac{1}{2p} \, \alpha_p^i \delta E_{ij}
\bar{\alpha}_p^j |0 \! >.
\end{equation}

\noindent Comparing this with Eqn.(~\ref{dncw0}) proves the result required.

\section{Showing that $\Delta L_n = \delta L_n$}
For any conformal theory the Virasoro operators can be determined once the
states corresponding to points in ${\cal P}(g,n)$ are known \cite{NCW}. Since
Eqn.(~\ref{mainncw}) determines the states of the perturbed conformal theory,
one can compute the perturbed Virasoro operators from it using the same method.
This was done in \cite{NCW} and written as equation (3.1.4) which we repeat
below
\begin{equation}
\Delta L_n = X_n = -\frac{1}{2 \pi i} \oint _{|z(Q)| =1}
d\bar{z}\,z^{n+1}\,\Phi_{z\bar{z}} (Q). \label{xn}
\end{equation}

\noindent For the particular space of backgrounds under consideration the
explicit expression for  $\Phi_{z\bar{z}}$ is given in Eqn.(~\ref{phi}).
Substituting for $\Phi_{z\bar{z}}$ in Eqn.(~\ref{xn}) and changing to the
variable $\theta$ where $z=re^{i\theta}$ with $r=1$ we get
\begin{eqnarray}
 X_n & = & \frac{i}{2 \pi i} \, \int_0^{2\pi} \! d\theta \, e^{-i\theta}
e^{i(n+1)\theta} \sum_p \alpha_p^i  e^{-i(p+1)\theta} \delta E_{ij} \sum_q
\bar{\alpha}_q^j e^{i(q+1)\theta},\\
& = & \frac{1}{2 \pi }  \sum_p \sum_q \int_0^{2\pi} d\theta  \,
e^{i(n+q-p)\theta} \alpha_p^i \delta E_{ij}\bar{\alpha}_q^j, \nonumber\\
& = & \sum_q \alpha_{(n+q)}^i \delta E_{ij} \bar{\alpha}_q^j. \label{dncwln}
\end{eqnarray}

Having computed $\Delta L_n$ we examine $\delta L_n$ which is defined to be
$\Gamma_{KZ} (E,E') L_n(E') - L_n(E)$. Here $\Gamma_{KZ} (E,E') L_n(E')$
denotes the operator obtained by transporting a virasoro operator from the
background $E' = E + \delta E$ to the background $E$ with the connection
$\Gamma_{KZ}$. Equation (2.28) of \cite{KZ} tells us that $\delta L_n = \sum_q
\alpha_{(n+q)}^i \delta E_{ij} \bar{\alpha}_q^j$. Comparison with
Eqn.(~\ref{dncwln}) shows that $\Delta L_n = \delta L_n$ as promised.

\section{Geometry on B}

In this section we discuss affine geometry on the space of backgrounds $B$
\cite{K}. The space $B$ is the homogeneous manifold $SO(d,d)/SO(d){\tt x}SO(d)$
and is hence equipped with a natural metric. This metric is the one invariant
under the action of $SO(d,d)$ on the space $B$. We have used a representation
for
the space B in terms of a $d$-dimensional matrix $E$. If
\begin{equation}
f = \left( \begin{array}{cc}
a & b\\
c & d
\end{array}
\right),
\end{equation}

\noindent denotes an element in $SO(d,d)$ where $a,b,c \& d$ are $d{\tt x}d$
matrices, then the action of this element on a matrix $E$ is given by
\begin{equation}
E' = f(E) \equiv (aE+b)(cE+d)^{-1} = \alpha \beta^{-1},
\end{equation}

\noindent where $\alpha = (aE + b)$ and $\beta = (cE +d)$. The metric invariant
under this action is given by
\begin{equation}
 g(C,D) =  Tr(G^{-1} C^T G^{-1} D) \label {metric},
\end{equation}
\noindent where $G \equiv E^S = \frac{1}{2}(E + E^T)$ and $C$ \& $D$ are
elements of $T_E B$. We think of $C$ and $D$ as matrices if no indices are
present (i.e. $C = C_{\mu \nu} \partial/\partial E_{\mu \nu}$). Otherwise we
use the notation $C^i$ or $D^j$ with the index $i$ or $j$ referring to the
pairs of indices of a matrix.  To check the invariance of this metric under a
transformation $g \in SO(d,d)$ we need to compute the push forward $f_*C$ and
$f_*D$ of the vectors $C$ and $D$ under this transformation and compute their
product at the new point. The push forward is
\begin{equation}
f_*C = (aC\beta^{-1} - \alpha\beta^{-1}cC\beta^{-1}),
\end{equation}

\noindent and from \cite{KZ} we have that
\begin{equation}
\beta^{-1} G'^{-1} (\beta^{-1})^T =  G^{-1} .
\end{equation}

\noindent This yields

\begin{eqnarray}
Tr (C_f^TG'^{-1}D_f^TG'^{-1})&=& Tr \left ( G^{-1} C^T (a -
\alpha\beta^{-1}c)^T \beta G^{-1} \beta^T (a - \alpha\beta^{-1}c) D
\right ),\nonumber \\
&=& Tr \left ( G^{-1} C^T (I - 2 G \beta^{-1}c)^T G^{-1} (I -
2G \beta^{-1}c)D \right ) \label {trmetric},
\end{eqnarray}

\noindent where in the last step we have used some of the identities \cite{KZ}
satisfied by the matrices $a,b,c\;\&\;d$. Invariance of the metric is
the requirement that the RHS of Eqn.(~\ref{trmetric}) is the same as
the RHS of Eqn.(~\ref{metric}). We will now use the fact that
$SO(d,d)$ is generated by the elements for which $c=0$ and the element
($a=d=0\;\&\;b=c=I$). For elements for $c=0$ quite clearly the
condition for isometry is satisfied. Let us examine the RHS of
Eqn.(~\ref{trmetric}) for the element ($a=d=0\;\&\;b=c=I$).
\begin{eqnarray}
Tr \left ( G^{-1} C^T (I - 2 G E^{-1})^T G^{-1} (I -
2G E^{-1})D \right ) & = & Tr(G^{-1} C^T G^{-1} D) \nonumber\\
&& - 2 Tr \left ( G^{-1} C^T \left( (E^{-1})^T + E^T \right ) D \right
) \nonumber\\
&& + 4 Tr \left ( G^{-1} C^T (E^{-1})^T G E^{-1} D \right )
\end{eqnarray}
\noindent Writing $G = \frac{1}{2}(E + E^T)$ in the last term yields
the desired result
\begin{equation}
Tr \left ( G^{-1} C^T (I - 2 G E^{-1})^T G^{-1} (I - 2G E^{-1})D \right ) =
Tr(G^{-1} C^T G^{-1} D).
\end{equation}

This invariant metric is the same as the Zamolodchikov metric \cite{GMR}. The
Zamolodchikov metric is constructed using the identification of $T_EB$ with
$M_E$. The BPZ product restricted to $M_E$ is moved via the identification to
$T_EB$ to obtain the Zamolodchikov metric. To verify that this metric is the
same as the invariant metric discussed earlier we compute the Zamolodchikov
metric
\begin{equation}
  <\! 0| \alpha_1^i C_{ij} \bar{\alpha}_1^j  \alpha_{-1}^i D_{ij}
\bar{\alpha}_{-1}^j |0\!>  =  Tr(G^{-1} C^T G^{-1} D) =  g(C,D).
\end{equation}

{}From metrics we now move to consider connections or in other words affine
geometry on $B$. The connection associated with the Zamolodchikov metric is
computed using the formula $\Gamma_{ijk} = \frac{1}{2}(\partial_j g_{ik} +
\partial_k g_{ij} - \partial_i g_{jk})$. We define $\Gamma^Z{[A,B,C]} =
\Gamma^Z_{ijk}A^iB^jC^k$ and have used the square brackets to indicate that
this it is not an invariant object. The formula $\delta G^{-1} = G^{-1}\:
\delta G \: G^{-1}$ tells us that $\partial_i G^{-1} A^i = G^{-1} \: A^S \:
G^{-1}$ where $A^S \equiv \frac{1}{2}(A + A^T)$. Using this result we find that
\begin{equation}
\partial_j g_{ik} A^i B^j C^k = Tr(A^T G^{-1} B^S G^{-1} C G^{-1}) + Tr(A^T
G^{-1} C G^{-1} B^S G^{-1}),
\end{equation}
\noindent with similar expressions for the other two terms. Using the cyclic
property of the trace we find that
\begin{equation}
\Gamma^Z{[A,B,C}] = -\frac{1}{2} \left ( Tr(A^T G^{-1} B G^{-1} C G^{-1}) +
Tr(A^T G^{-1} C G^{-1} B G^{-1}) \right ). \label {gammaz}
\end{equation}

Having computed the connection associated with the Zamolodchikov metric
consider the connection $P\Gamma_{KZ}$ obtained by projecting $\Gamma_{KZ}$
onto the subspace $M_E$. The connection $\Gamma_{KZ}$ is an object with index
structure $\Gamma^X_{Yi}$ where $X$ and $Y$ index the state space of a theory
$H_E$ and $i$ indexes the tangent space $T_E B$. An alternative to the index
notation is to use the number $\Gamma_{KZ}{[\alpha, \beta, C}]$ obtained on
choosing frames on the fibers and choosing elements $\alpha$ and $\beta$ in the
fiber $H_E$ and an element $C$ in the tangent space $T_E B$. We want to compute
this number for the case in which $\alpha$ and $\beta$ are in $M_E$ and are
associated with the matrices A and B respectively. So we begin with the state
$\alpha_{-1}^i B_{ij} \bar{\alpha}_{-1}^j |0\!>_{E'}$  in $M_{E'}$ and apply
the prescription of Eqn.(~\ref{conkz}). We find
\begin{eqnarray}
\Gamma_{KZ}{[\cdot,B,C]} & = & - \frac{1}{2} G^{il} C_{kl}\alpha_{-1}^k B_{ij}
\bar{\alpha}_{-1}^j |0\!> - \frac{1}{2} \alpha_{-1}^i B_{ij} G^{jl}
C_{lk}\bar{\alpha}_{-1}^k |0\!> \nonumber\\
& & - \frac{1}{2} G^{il} C_{lk}\bar{\alpha}_{-1}^k B_{ij} \bar{\alpha}_{-1}^j
|0\!> - \frac{1}{2} \alpha_{-1}^i B_{ij} G^{jl} C_{kl}\alpha_{-1}^k
|0\!>\nonumber\\
& & + \alpha_{-1}^i B_{ij} \bar{\alpha}_{-1}^j {\cal B} |0\!>  \label{trans}
\end{eqnarray}

\noindent Now to compute $P\Gamma_{KZ}$ project this state onto $M_E$. To do so
just take the BPZ product of this with the state in $M_E$ determined by the
matrix A, since this is the state which we will ultimately use as the first
entry in $P\Gamma_{KZ}$. The last three terms of Eqn.(~\ref{trans}) yield zero
on taking the BPZ product and so we get
\begin{equation}
P\Gamma_{KZ}{[A,B,C]} = <\!0| \alpha_1^p A_{pq} \bar{\alpha}_1^q
\left(-\frac{1}{2} G^{il} C_{kl}\alpha_{-1}^k B_{ij} \bar{\alpha}_{-1}^j |0\!>
\;\;-\;\;\frac{1}{2} \alpha_{-1}^i B_{ij} G^{jl} C_{lk}\bar{\alpha}_{-1}^k
|0\!> \right ).
\end{equation}

\noindent Evaluating this we get
\begin{equation}
P\Gamma_{KZ}{[A,B,C]} = -\frac{1}{2} \left ( Tr(A^T G^{-1} B G^{-1} C G^{-1}) +
Tr(A^T G^{-1} C G^{-1} B G^{-1})\right ). \label {pkz}
\end{equation}

\noindent Comparison of Eqn.(~\ref{gammaz}) with Eqn.(~\ref{pkz}) establishes
that $P\Gamma_{KZ} = \Gamma^Z$.

\section{Conclusions and Questions}

The chief result of this paper is that for the space of theories considered
there is a connection satisfying the condition for first order equivalence.
This condition establishes the equivalence of deformed theories with those in a
neighbouring background to first order in the perturbation parameter
$\epsilon$. Such a connection could be relevant to the current efforts to
formulate a background independent string field theory where one is attempting
to write a lagrangian for the ``space of theories" \cite{W}. Presumably in such
a formulation the tangent space at a point is the state space of the theory at
that point. We expect that in such a formulation there is a method of deforming
theories within a single state space.  The possibility of such deformation
suggests a role for a connection on the ``space of theories" to establish the
equivalence of a deformed theory with a neighbouring one.

That the connection $\Gamma_{KZ}$ satisfies the condition for first order
equivalence is noteworthy. In fact it is remarkable considering that the
connection was defined in terms of ``universal coordinates" \cite{KZ}, a
structure independent of the definition of CFT's. In addition the connection
$P\Gamma_{KZ}$ induced on the space of backgrounds by $\Gamma_{KZ}$ is the same
as the connection that arises from the Zamolodchikov metric. These two
``coincidences" suggests that there is a deeper relationship between
``universal coordinates" and conformal/string theory which needs to be
elucidated.  One direction of investigation would be to seek an analogue of
``universal coordinates" in more general conformal/string theories.

This paper however does not address the question of existence of actual (i.e
finite as opposed to infinitesimally) deformations. The second and higher order
corrections to Eqn.(~\ref{mainncw}) could perhaps be defined in a manner that
respects the sewing axiom, but to our knowledge this issue still needs
investigation. We have also not investigated the possiblity of defining actual
maps to establish equivalence of theories, but have rather dealt only with
connections. We need to understand whether solutions to the equation of
parallel transport asociated to the connection exist and whether the maps so
defined establish the equivalence of deformed theories to those at a finitely
separated background.

Some of the issues mentioned in these concluding remarks have been
clarified in a recent paper \cite{RSZ}.

\paragraph{Acknowledgments}
I would like to thank B.Zwiebach for suggestions during the work. I am grateful
to S.Mathur for useful conversations.

\section*{Appendix: $\Delta |p\!> = \delta |p\!>$ for p a contact interaction}

As explained earlier a contact interaction is one where the $|\xi_i(z)| \leq 1$
discs for each puncture cover the entire surface exactly. It is clear that
$\Delta |p\!> = 0$ in this case because there is no area exterior to the discs
on which the integral of Eqn.(~\ref{mainncw}) can be performed.

We want to now check that $\delta |p\!> = 0$ for contact interactions. In
\cite{KZ} the expression for the the tensor $|p\!>$ was given. We recall the
relevant part of this expression. For the full expression see Eqn.(3.24) of
\cite{KZ}.
\begin{equation}
|p\!> = (...) \cdot \exp \left (\frac{1}{2}\sum_{r,s} \sum_{n,m \leq 0}
N_{nm}^{rs} \alpha_n^{i(r)}(E)G_{ij}\alpha_m^{i(r)}(E) + \frac{1}{2}\sum_{r,s}
\sum_{n,m \leq 0} \bar{N}_{nm}^{rs}
\bar{\alpha}_n^{i(r)}(E)G_{ij}\bar{\alpha}_m^{i(r)}(E)\right ) |0\!>_E.
\end{equation}

\noindent The conditions that the Neumann coefficients of a vertex must satisfy
so that $\delta |p\!> = 0$, were derived in \cite{KZ} (See equations (3.48)). A
minor correction needs to be made to those identities since they were derived
assuming that the Neumann coefficients for the antiholomorphic sector are the
same as those of the holomorphic sector. In fact the coefficients for the
antiholomorphic sector are the complex conjugates of those in the holomorphic
sector and hence the correction. We simply list the corrected conditions below.
\begin{eqnarray}
\bar{N}^{rs}_{m_1n} n N^{st}_{nm_2} & = & \frac{1}{m_1} \delta_{m_1m_2}
\delta^{rt}, \label{one}\\
\bar{N}^{rs}_{0n} n N^{st}_{nm} & = & - N^{rt}_{0m}, \label{two}\\
\bar{N}^{rs}_{0n} n N^{st}_{n0} & = & -(N^{rt}_{00} + \bar{N}^{rt}_{00}).
\label{three}
\end{eqnarray}

The Neumann coefficients have an integral representation \cite{LPP}. We will
use these integral representations to verify that Eqns.(~\ref{one}),
(~\ref{two}) \& (~\ref{three}) above are satisfied for any contact interaction.
The integral representations involve the functions $h_i(z)$. The $h_i(z)$
define the local coordinates around the punctures by mapping a neighbourhood of
the origin in the z plane to a neighbourhood of the puncture $h_i(0)$. By
changing variables to the inverse of these functions $\xi_i(u)$ we write them
in the form we will need later.
\begin{equation}
N^{rs}_{00} = \left\{\begin{array}{ll}
log |h'_{r}(0)|  & \mbox {for $r=s$} \label{integ1}\\
log |h_r(0) - h_s(0)| & \mbox {for $r \neq s$}
\end{array}
\right.
\end{equation}
\begin{eqnarray}
N^{rs}_{0m} & = & \frac{1}{m} \oint \frac{dw}{2 \pi i} w^{-m} (h'_s(w))
\frac{-1}{(h_r(0) - h_s(w))}\nonumber\\
& =  & \mbox{  }\oint \frac{du}{2 \pi i} \xi_s(u)^{-m} \frac{1}{(h_r(0) - u)}\\
N^{rs}_{nm} & = & \frac{1}{n} \oint \frac{dw}{2 \pi i} w^{-n} (h'_r(w))
\frac{1}{m} \oint \frac{dz}{2 \pi i} z^{-m} (h'_s(z)) \frac{-1}{(h_r(w) -
h_s(z))^2}\nonumber\\
& = & \frac{1}{n}\oint \frac{dy}{2 \pi i} \xi_r(y)^{n} \oint\frac{du}{2 \pi i}
\xi_s(u)^{-m} \frac{1}{(y - u)^2}
\end{eqnarray}

To verify Eqns.(~\ref{one}),(~\ref{two}) \& (~\ref{three}) we will evaluate the
left hand side of each in turn by carrying out the sums and the integrals
involved. Before we begin we note some of the common features that will arise
in manipulating the integrals. The first point to note is that we will not
consider a point in  $p \in {\cal P}(g,n)$ corresponding to a contact
interaction directly. We will scale all the local coordinates $\xi$ by the
factor $(1+\epsilon)$ to define the new local coordinate $\xi^{\epsilon}(z)$ .
The contour $|\xi^{\epsilon}(z)| = 1$ lies inside the $|\xi (z)| = 1$ contour
leaving some area on the surface outside of the new circles where we know that
the local coordinate $\xi^{\epsilon}(z)$ is well defined and analytic. This
area will be necessary in the course of the proofs below. After doing all the
manipulations however we will find that the limit $\epsilon \rightarrow 0$
exists, yielding the identities we seek for contact interactions. In dealing
with each of the identities we will, on manipulating the left hand side get the
desired right hand side and an additional term that goes to zero in the limit
$\epsilon \rightarrow 0$. The mechanism for this additional term going to zero
will be the same in all cases; namely ``pairwise cancellation", the meaning of
which will be explained later. We will however not write out the process of
scaling and then take the limit explicitly since it will be fairly obvious how
our arguments below can be made rigorous.

The second issue is that the types of manipulations fall into three categories.
The kind of quantity one starts out with is a sum of multiple integrals and so
the manipulations are:  1) Deforming contours and breaking one contour into two
(eg when we move a contour across a singularity) 2) Algebraic operations such
as summing up terms/rearranging the sums and integration by parts - the result
of either of these is to change the integrands we will deal with 3) actually
carry out the integration around a contour using the residue method resulting
in an integral with one less variable(contour). As regards 3) we will always do
first the integration in the variable $v$, then in $u$, then if necessary in
$y$ and $z$ where the role of variables $u,v,y \& z$ is defined below. (There
is a minor modification to this order of evaluation in proving the second part
of the last identity).

Lastly Fig. 2 indicates the representative location of the
contours.  In each case the $u$ contour will be the $|\xi_s^{\epsilon} (u)| = 1
$ contour. The $v$ contour will start out just outside of the $u$ contour but
within the contour $|\xi_s (v)| = 1$, and then be subject to deformations. The
need for this arises as follows.  We will need in the proof to locate the $u$
contour as the  $|\xi_s^{\epsilon} (u)| = 1 $ curve. In the very first step of
each of the three proofs one will do the summation over n. This sum is yields $
\log (1 - \frac{1}{\bar{\xi}_s^{\epsilon}(u) \xi_s^{\epsilon}(v)})$ which is
convergent for $|\xi_s^{\epsilon} (u)| = 1$ only if $v$ satisfies
$\xi_s^{\epsilon}(v) > 1$. Since all reference to $\epsilon$ is dropped below,
these remarks on initial location of contours should be borne in mind.

\subsection*{First condition: $\sum_{s} \sum_{n=1}^{\infty}
\bar{N}^{rs}_{m_1,n} n N^{st}_{n,m_2}  = \delta^{rt} \frac{1}{m_1}
\delta_{m_1m_2}$}

To verify this condition we evaluate the left hand side using the integral
representation. This gives a sum over integrals of four variables $u,v,y \& z$
with $y$ and $u$ associated with the first Neumann coefficient and $z$ and $v$
being associated with the second. The left hand side then is
\begin{equation}
\sum_{s} \sum_{n=1}^{\infty} \bar{N}^{rs}_{m_1n} n N^{st}_{nm_2} =
\frac{1}{m_1} \oint \frac{\bar{dy}}{2 \pi i} \bar{\xi}_r(y)^{-m_1}
\frac{1}{m_2} \oint \frac{dz}{2 \pi i} \xi_t(z)^{-m_2}  F(\bar{y},z),
\label{onef}
\end{equation}

\noindent where
\begin{equation}
F(\bar{y},z) = \sum_{s} \oint \frac{\bar{du}}{2 \pi i} \frac{1}{(\bar{y} -
\bar{u})^2} \oint \frac{dv}{2 \pi i}  \frac{1}{(v - z)^2} \log (1 -
\frac{1}{\bar{\xi}_s(u) \xi_s(v)})
\end{equation}

\noindent We first evaluate $F(\bar{y},z)$. To do so we integrate by parts in
the variable $v$. Since the $v$ contour encloses the branch cut singularity in
the logarithm (see Fig.2) we have one term that is the integral of the total
derivative of a single valued function. This yields 0. So we obtain
\begin{equation}
F(\bar{y},z) = \sum_{s} \oint\frac{\bar{du}}{2 \pi i} \frac{1}{(\bar{y} -
\bar{u})^2} \oint \frac{dv}{2 \pi i}  \frac{1}{(v - z)}
\frac{\xi_s'(v)}{\xi_s(v)(1 - \bar{\xi}_s(u)\xi_s(v))}
\end{equation}
\noindent We evaluate this integral by finding the residues at the poles. There
are poles at $v = h_s(0)$ and  $v=u$, for all s. For $s=t$ there is an
additional pole that is encircled at $v=z$.  Doing these integrals we find
\setcounter{dump}{\value{equation}}
\setcounter{equation}{0}
\renewcommand{\theequation}{\Roman{equation}}
\begin{eqnarray}
F(\bar{y},z) & = & \sum_{s} \oint\frac{\bar{du}}{2 \pi i}  \frac{1}{(\bar{y} -
\bar{u})^2} \frac{1}{(z-h_s(0))} \\
& & \mbox{} + \sum_{s} \oint\frac{\bar{du}}{2 \pi i}  \frac{1}{(\bar{y} -
\bar{u})^2}\frac{1}{(z - u)}\\
& & \mbox{} + \oint \frac{du}{2 \pi i} \frac{1}{(\bar{y} - \bar{u})^2}
\frac{\xi_t'(z)}{\xi_t(z) (\bar{\xi}_t(u) \xi_t(z) - 1)}.
\end{eqnarray}
\setcounter{equation}{\value{dump}}
\renewcommand{\theequation}{\arabic{equation}}

\noindent We evaluate each of the terms by doing the $u$ integration. We find
\begin{eqnarray*}
(I) & = & 0  \mbox{ because the residue of  $\frac{1}{(\bar{y} - \bar{u})^2}$
at $u=y$ is 0.}\\
(II) & = & 0 \mbox{ because of pairwise cancellation}.\\
(III) & = & \delta^{rt} \frac{\xi_t'(z)\bar{\xi}_t'(y)}{(1 -
\xi_t(z)\bar{\xi}_t(y))^2}.
\end{eqnarray*}

\noindent Plugging expression (III) for $F(\bar{y},z)$ into Eqn.(~\ref{onef})
we get
\begin{equation}
\sum_{s} \sum_{n=1}^{\infty} N^{rs}_{m_1,n} n N^{st}_{n,m_2}
= \delta^{rt}  \frac{1}{m_1} \oint \frac{dy}{2 \pi i} \xi_t(y)^{-m_1}
\frac{1}{m_2} \oint \frac{dz}{2 \pi i} \xi_t(z)^{-m_2}
\frac{\xi_t'(z)\bar{\xi}_t'(y)}{(1 -  \xi_t(z)\bar{\xi}_t(y))^2}
\end{equation}

\noindent We write $\frac{1}{(1 -  \xi_t(z)\bar{\xi}_t(y))^2} =
\sum_{n=0}^{\infty} (n+1) (\xi_t(z)\bar{\xi}_t(y))^n$. We then change variables
from $y$ and $z$ to $\bar{\xi}_t(y)$ and $\xi_t(z)$. We see that $m_1$ must
equal $m_2$ for a non zero answer. We finally get the result
\begin{equation}
\sum_{s} \sum_{n=1}^{\infty} \bar{N}^{rs}_{m_1,n} n N^{st}_{n,m_2}  =
\delta^{rt} \frac{1}{m_1} \delta_{m_1m_2}.
\end{equation}

\subsection*{Second condition: $\sum_{s}\sum_{n=1}^{\infty} N^{rs}_{0n} n
\bar{N}^{st}_{nm}= - \bar{N}^{rt}_{0m}$}

As in the earlier case we evaluate the left hand side using the integral
representation. In this case there are only three variables $u,v,\& z$ with $u$
being associated with the first Neumann coefficient and $z$ and $v$ being
associated with the second. We find
\begin{equation}
\sum_{s} \sum_{n=1}^{\infty} N^{rs}_{0n} n \bar{N}^{st}_{nm} = \frac{1}{m}
\oint \frac{\bar{dz}}{2 \pi i}\bar{\xi}_t(z)^{-m}  G(z), \label{g}
\end{equation}
\noindent where
\begin{equation}
G(\bar{z}) = \sum_{s} \oint \frac{du}{2 \pi i}  \frac{1}{(h_r(0) - u)}  \oint
\frac{\bar{dv}}{2 \pi i} \frac{1}{(\bar{v} - \bar{z})^2}\log (1 -
\frac{1}{\bar{\xi}_s(v) \xi_s(u)}).
\end{equation}
\noindent We first evaluate $G(\bar{z})$. To do so we integrate by parts in the
variable v to obtain two terms. The first is the integral of the total
derivative of a single valued function - single valued since the $v$ contour
encloses the branch cut singularity of the logarithm (see Fig. 2). This term
yields 0. The other term yields
\begin{equation}
G(\bar{z}) = \sum_{s} \oint \frac{du}{2 \pi i}  \frac{1}{(h_r(0) - u)}  \oint
\frac{\bar{dv}}{2 \pi i} \frac{1}{(\bar{v} - \bar{z})}
\frac{\xi_s'(v)}{\xi_s(v)(1 - \bar{\xi}_s(u)\xi_s(v))}.
\end{equation}
\noindent Now we perform the $v$ integration by finding the residues at the
poles. There are poles at $v = h_s(0)$ and  $v=u$, for all s. For $s=t$ there
is an additional pole that is encircled at $v=z$. Doing these integrals we find
\setcounter{dump}{\value{equation}}
\setcounter{equation}{0}
\renewcommand{\theequation}{\Roman{equation}}
\begin{eqnarray}
G(\bar{z}) & = & \sum_{s} \oint\frac{du}{2 \pi i} \frac{1}{(h_r(0) - u)}
\frac{1}{(h_s(0) - \bar{z})}\\
& & \mbox{} + \sum_{s} \oint\frac{du}{2 \pi i} \frac{1}{(h_r(0) - u)}
\frac{1}{(\bar{u} - \bar{z})}\\
& & \mbox{} +  \oint_{h_t(0)} \frac{du}{2 \pi i}  \frac{1}{(h_r(0) - u)}
\frac{\bar{\xi}_t'(z)}{\bar{\xi}_t(z) (\xi_t(u) \bar{\xi}_t(z) - 1)}.
\end{eqnarray}
\setcounter{equation}{\value{dump}}
\renewcommand{\theequation}{\arabic{equation}}

\noindent Evaluating each of the terms in order we find
\begin{eqnarray*}
(I) & = & \frac{1}{(\bar{h}_r(0) - \bar{z})} \;\; \mbox{since only contribution
is from the pole $\frac{1}{(h_r(0) - u)}$ when $s=t$}.\\
(II) & = & 0 \mbox{ by pairwise cancellation.}\\
(III) & = & \delta ^{rt}  \frac{\bar{\xi}_t'(z)}{\bar{\xi}_t(z)} \mbox {since
the $u$ contour encircles a pole only if $t=r$}
\end{eqnarray*}

\noindent Now we put back (I) and (III) back into Eqn.(~\ref{g}) to get
respectively
\begin{equation}
\frac{1}{m} \oint \frac{\bar{dz}}{2 \pi i}\bar{\xi}_t(z)^{-m} \;\;(I) =
\frac{1}{m} \oint \frac{\bar{dz}}{2 \pi i}\bar{\xi}_t(z)^{-m}
\frac{1}{(\bar{h}_r(0) - \bar{z})} = - \bar{N}^{rt}_{0m}
\end{equation}
\begin{equation}
\frac{1}{m} \oint \frac{\bar{dz}}{2 \pi i}\bar{\xi}_t(z)^{-m}\;(III) =
\frac{1}{m} \oint \frac{\bar{dz}}{2 \pi i}\bar{\xi}_t(z)^{-m} \delta ^{rt}
\frac{\bar{\xi}_t'(z)}{\bar{\xi}_t(z)} = \delta ^{rt} \frac{1}{m} \oint
\frac{\bar{dw}}{2 \pi i} w ^{-m-1} = 0
\end{equation}
\noindent So finally one has the result sought after
\begin{equation}
\sum_{s} \sum_{n=1}^{\infty} N^{rs}_{0n} n \bar{N}^{st}_{nm}=
-\bar{N}^{rt}_{0m}.
\end{equation}

\subsection*{Third condition: $\sum_{s}\sum_{n}\bar{N}^{rs}_{0n} n N^{st}_{n0}
=  -(N^{rt}_{00} + \bar{N}^{rt}_{00})$}

To evaluate the left hand side for this condition we will need to consider
separately the cases $r \neq t$ and $r=t$. This is needed because the nature of
the integrals that occur differ for these two cases. The separate treatment is
to be expected since the right hand side (see Eqn ~\ref{integ1} has a form that
depends on whether $r=t$ or not.  The only variables of integration here are
$u$ \& $v$ with $u$ being associated with the first Neumann coefficient and
$v$, the second.  First consider the case of $r \neq t$

\subsubsection*{\underline{The case $r \neq t$}}

We begin by writing down the left hand side using the integral representation
of the Neumann functions. This leads to
\begin{equation}
\sum_{s}\sum_{n}\bar{N}^{rs}_{0n} n N^{st}_{n0} = \sum_{s}\oint_{h_s(0)}
\frac{\bar{du}}{2 \pi i} \frac{1}{(\bar{h}_r(0) - \bar{u})}  \oint_{h_s(0)}
\frac{dv}{2 \pi i} \frac{1}{(v-h_t(0))}\log (1 - \frac{1}{\bar{\xi}_s(u)
\xi_s(v)}).
\end{equation}

\noindent We integrate by parts in the variable v. In doing so we will have to
treat the cases $s=t$ and $s \neq t$ differently. In the latter case the
function $\frac{1}{(v-h_t(0))}$ is analytic in the contour and so its integral
$\log(v-h_t(0))$ can be choosen to be single valued. In the former case its
integral $\log(v-h_t(0))$ is not single valued and so the usual integral of a
total derivative is not zero. So for $s=t$ let $V$ be the value of $v$ where
the contour in the v plane intersects the branch cut in the function
$\log(v-h_t(0))$. We use $V_+$ and $V_-$ to refer respectively to the value of
$v$ just above and just below the branch cut at the point $V$. Then we find
that the left hand side
\begin{eqnarray}
 & = & \sum_{s \neq t} \oint_{h_s(0)} \frac{\bar{du}}{2 \pi i}
\frac{1}{(\bar{h}_r(0) - \bar{u})}  \oint_{h_s(0)} \frac{dv}{2 \pi i}
\log(v-h_t(0)) \frac { \xi_s'(v)} {\xi_s(v) (\bar{\xi}_s(u) \xi_s(v) -
1)}\nonumber \\
& \mbox{  } + & \oint_{h_t(0)} \frac{\bar{du}}{2 \pi i} \frac{1}{(\bar{h}_r(0)
- \bar{u})} \int_{V_-}^{V_+} \frac{dv}{2 \pi i} \log(v-h_t(0)) \frac {
\xi_t'(v)} {\xi_t(v) (\bar{\xi}_t(u) \xi_t(v) - 1)}\nonumber \\
& \mbox{  } + & \oint_{h_t(0)} \frac{\bar{du}}{2 \pi i} \frac{1}{(\bar{h}_r(0)
- \bar{u})} \log (1 - \frac{1}{\bar{\xi}_t(u) \xi_t(V)}).
\end{eqnarray}

\noindent In the first term we get contributions in the $v$ plane from the
poles at $v = h_s(0)$ and $v=u$. In the second term we move the contour in the
$v$ plane across the pole at $v=u$. We group the contribution from the pole at
$u$ with similar terms arising in the case $s \neq t$. The term that remains
is zero, since on changing the order of integration we
find that the contour in the u plane encloses no singularities. In the third
term we integrate by parts in the variable $u$. Doing all of this we find
\setcounter{dump}{\value{equation}}
\setcounter{equation}{0}
\renewcommand{\theequation}{\Roman{equation}}
\begin{eqnarray}
\sum_{s}\sum_{n}\bar{N}^{rs}_{0n} n N^{st}_{n0} & = & \sum_{s}\oint_{h_s(0)}
\frac{\bar{du}}{2 \pi i} \frac{1}{(\bar{h}_r(0) - \bar{u})} \log(u-h_t(0))\\
&\mbox{  } + & \sum_{s \neq t} \oint_{h_s(0)} \frac{\bar{du}}{2 \pi i}
\frac{1}{(\bar{h}_r(0) - \bar{u})}  \log(h_s(0)-h_t(0))\\
&\mbox{  } + & \oint_{h_t(0)} \frac{\bar{du}}{2 \pi i} \log (\bar{h}_r(0) -
\bar{u}) \frac {\bar{\xi}_t'(u)} {\bar{\xi}_t(u) (\bar{\xi}_t(u) \xi_t(V) -
1)}.
\end{eqnarray}
\setcounter{equation}{\value{dump}}
\renewcommand{\theequation}{\arabic{equation}}
\noindent Examining each term in turn yields
\begin{eqnarray*}
(I) & = & 0 \mbox{ by pairwise cancellation.}\\
(II) & = & - \log(h_r(0) - h_t(0)) \mbox{ since only the $s=r$ term
contributes.}\\
(III) & = & - \log(\bar{h}_r(0) - \bar{h}_t(0)) \mbox{ since the only pole
enclosed is at $u=h_t(0)$}
\end{eqnarray*}

\noindent So we get finally what was to be shown for the case $r\neq t$
\begin{eqnarray}
\sum_{s}\sum_{n}\bar{N}^{rs}_{0n} n N^{st}_{n0} & = & - \log(h_r(0) - h_t(0)) -
\log(\bar{h}_r(0) - \bar{h}_t(0)) \nonumber \\
 & = & -2 \:\log|h_r(0) - h_s(0)| =  -(N^{rt}_{00} + \bar{N}^{rt}_{00}).
\end{eqnarray}

\subsubsection*{\underline{The case $r=t$}}

Just as in the earlier case we use the integral representation to write the
left hand side. We find
\begin{equation}
\sum_{s}\sum_{n}\bar{N}^{ts}_{0n} n N^{st}_{n0} = \sum_{s}\oint_{h_s(0)}
\frac{\bar{du}}{2 \pi i} \frac{1}{(\bar{h}_t(0) - \bar{u})}  \oint_{h_s(0)}
\frac{dv}{2 \pi i} \frac{1}{(v-h_t(0))}\log (1 - \frac{1}{\bar{\xi}_s(u)
\xi_s(v)})
\end{equation}

\noindent We now go through the steps of the earlier subsection. Integrate by
parts in the variable $v$. Group together the terms for $s \neq t$. For $s=t$
the integration by parts picks up an additional term. For $s=t$ let $V$ be the
value of $v$ where the contour in the $v$ plane intersects the branch cut in
the function $\log(v-h_t(0))$. Then we find that the left hand side
\begin{eqnarray}
& = & \sum_{s \neq t} \oint_{h_s(0)} \frac{\bar{du}}{2 \pi i}
\frac{1}{(\bar{h}_t(0) - \bar{u})}  \oint_{h_s(0)} \frac{dv}{2 \pi i}
\log(v-h_t(0)) \frac { \xi_s'(v)} {\xi_s(v) (\bar{\xi}_s(u) \xi_s(v) - 1)}
\nonumber \\
& \mbox{  } + & \oint_{h_t(0)} \frac{\bar{du}}{2 \pi i} \frac{1}{(\bar{h}_t(0)
- \bar{u})} \int_{V_-}^{V_+} \frac{dv}{2 \pi i} \log(v-h_t(0)) \frac {
\xi_t'(v)} {\xi_t(v) (\bar{\xi}_t(u) \xi_t(v) - 1)} \nonumber \\
& \mbox{  } + & \oint_{h_t(0)} \frac{\bar{du}}{2 \pi i} \frac{1}{(\bar{h}_t(0)
- \bar{u})} \log (1 - \frac{1}{\bar{\xi}_t(u) \xi_t(V)}).
\end{eqnarray}

\noindent In the first term we get contributions in the $v$ plane from the
poles at $v=h_s(0)$ and $v=u$. In the second term we move the contour in the
$v$ plane across the pole at $v=u$ and group as before the contribution from
the pole with similar terms arising in the first term. On the modified second
term (i.e with contour excluding the pole at $u$), we perform integration by
parts. In the third term also we integrate by parts in the variable $u$. In
both of these integration by parts we find that $\log (\bar{h}_t(0) - \bar{u})$
the integral of $\frac{1}{(\bar{h}_t(0) - \bar{u})}$ is not single valued on
the contour. Let $U$ denote the point at which the contour intersects the
branch cut in $\log (\bar{h}_t(0) - \bar{u})$. Also let $U_+$ and $U_-$ denote
respectively the points just above and just below the intersection point $U$.
With this notation we find on performing the operations above that the left
hand side
\setcounter{dump}{\value{equation}}
\setcounter{equation}{0}
\renewcommand{\theequation}{\Roman{equation}}
\begin{eqnarray}
   &=&\sum_{s} \oint_{h_s(0)} \frac{\bar{du}}{2 \pi i} \frac{1}{(\bar{h}_t(0) -
\bar{u})} \log(u-h_t(0))\\
   &+&\sum_{s \neq t} \oint_{h_s(0)} \frac{\bar{du}}{2 \pi i}
\frac{1}{(\bar{h}_t(0) - \bar{u})}  \log(h_s(0)-h_t(0))\\
   &+&\int_{U_-}^{U_+}\!\!\frac{\bar{du}}{2 \pi i} \!\log (\bar{h}_t(0) -
\bar{u}) \!\!\int_{V_-}^{V_+} \frac{dv}{2 \pi i} \log(v-h_t(0))\!\frac
{\xi_t'(v) \bar{\xi}_t'(u)} {(\bar{\xi}_t(u) \xi_t(v) - 1)^2}\\
   &+&\int_{U_-}^{U_+}\frac{\bar{du}}{2 \pi i} \log (\bar{h}_t(0) - \bar{u})
\frac {\bar{\xi}_t'(u)} {\bar{\xi}_t(u) (\bar{\xi}_t(u) \xi_t(V) - 1)}\\
   &+& \int_{V_-}^{V_+} \frac{dv}{2 \pi i} \log(v-h_t(0)) \frac { \xi_t'(v)}
{\xi_t(v) (\bar{\xi}_t(U) \xi_t(v) - 1)}\\
   &+&\log (1 - \frac{1}{\bar{\xi}_s(U) \xi_s(V)}).
\end{eqnarray}
\setcounter{equation}{\value{dump}}
\renewcommand{\theequation}{\arabic{equation}}

\noindent Examining each term in turn yields
\begin{eqnarray*}
(I) & = & 0 \mbox{ by pairwise cancellation.}\\
(II) & = & 0 \mbox{ since there are no singularities in the u plane for $s \neq
t$.}
\end{eqnarray*}

\noindent In terms (III) through (VI) we will shrink the contours in the u and
v planes to tightly enclose the singularities at $h_t(0)$ in both planes. This
means that U and V go to $h_t(0)$. In this limit we find
\begin{eqnarray*}
(III) & = & 0,\\
(IV)  & = & -\log(\bar{U} - \bar{h}_t(0)),\\
(V)   & = & - \log(V - h_t(0)),\\
(VI)  & = & \log {[(\bar{U} - \bar{h}_t(0))(V - h_t(0))
\xi_t'(h_t(0))\bar{\xi}_t'(h_t(0))]}.
\end{eqnarray*}

\noindent Writing the logarithm of the product in (VI) as a sum of logarithms
we find that (IV) and (V) are canceled by the first two logarithms. This
ensures that the sum of terms (III), (IV), (V) and (VI) is independent of the
value of $U$ and $V$ in the limit that they go to $h_t(0)$. So we finally get
\begin{equation}
\sum_{s}\sum_{n}\bar{N}^{ts}_{0n} n N^{st}_{n0} = \log
\xi_t'(h_t(0))\bar{\xi}_t'(h_t(0)) = - 2 \log | h_t'(0)| = -(N^{tt}_{00} +
\bar{N}^{tt}_{00}).
\end{equation}

\noindent In the last step we have used the fact that $\xi_t'(h_t(0))$ is
$\frac{1}{h_t'(0)}$, a consequence of the fact that $\xi$ and $h$ are inverses
of each other.

With the verification of this last identity we have proven that $\Delta |p\!>_E
= \delta |p\!>_E$ for $p$ a contact interaction.

\section*{Figure Caption}

\underline{Fig.2:} The figure shows a part of the Riemann sphere with the
$z=1$ contours of the local coordinates forming the grid for a contact
interaction. The representative initial location of the contours
is indicated. This figure applies to the first identity.
For the other identities one or both of the $y$ and $z$ contours will be
absent.

\end{document}